# Magnetotransport and lateral confinement in an InSe van der Waals Heterostructure

Yongjin Lee [1], Riccardo Pisoni [1], Hiske Overweg [1], Marius Eich [1], Peter Rickhaus [1], Amalia Patanè [2], Zakhar R. Kudrynskyi[2], Zakhar. D. Kovalyuk [3], Roman Gorbachev [4], Kenji Watanabe [5], Takashi Taniguchi [5], Thomas Ihn[1], and Klaus Ensslin[1]

[1]*Department of Physics, ETH Zurich, Otto-Stern-Weg 1, 8093 Zurich, Switzerland,*
[2]*School of Physics and Astronomy, The University of Nottingham, NG7 2RD, Nottingham, UK*
[3]*Institute for Problems of Materials Science, The National Academy of Sciences of Ukraine, Chernivtsi Branch, Chernivtsi 58001, Ukraine*
[4]*National Graphene Institute, University of Manchester, Manchester, M13 9PL, UK*
[5]*National Institute for Material science, 1-1 Namiki, Tsukuba 305-0044, Japan*

*E-mail : yongjin@phys.ethz.ch*

**Abstract**

**In the last six years, Indium selenide (InSe) has appeared as a new van der Waals heterostructure platform which has been extensively studied due to its unique electronic and optical properties. Such as transition metal dichalcogenides (TMDCs), the considerable bandgap and high electron mobility can provide a potential optoelectronic application. Here we present low-temperature transport measurements on a few-layer InSe van der Waals heterostructure with graphene-gated contacts. For high magnetic fields, we observe magnetoresistance minima at even filling factors related to two-fold spin degeneracy. By electrostatic gating with negatively biased split gates, a one-dimensional channel is realized. Close to pinch-off, transport through the constriction is dominated by localized states with charging energies ranging from 2 to 5 meV. This work opens new possibility to explore the low-dimensional physics including quantum point contact and quantum dot.**

**Introduction**

Indium selenide (InSe) is a novel 2D semiconductor that belongs to the family of post-transition metal chalcogenides, which have gained significant attention due to their unique electronic properties such as large variation of the bandgap as a function of number of layers.[1–5] The property that the bandgap of InSe increases significantly with decreasing number of layers open a possibility to facilitate bandgap engineering technology. Furthermore, unlike transition metal dichalcogenides (TMDCs) that have a direct band-gap only in films with one or two layers[6,7], InSe has a direct-band gap over a wide range of layer numbers.[8] Even when the crystal becomes an indirect band gap semiconductor, it remains optically active, enabling the fabrication of devices that are highly photosensitive over an extended wavelength range from the near-infrared to the visible.[9–14] The valence band of few-layer InSe exhibits an almost flat and slightly inverted "Mexican hat" near the $\Gamma$ point which could provide an opportunity to get strong correlations in hole-doped material such as superconductivity and tuneable magnetism.[15–18]

Most importantly, the in-plane electron masses are significantly lower than in TMDCs.[19] In recent experiments, the mobility of few-layer InSe has surpassed ~1,000 $cm^2$/Vs and ~12,000 $cm^2$/Vs at room temperature and 4.2 K, respectively.[1] These extraordinary electrical properties provide an excellent platform for 2D electronics and applications of InSe in low-dimensional devices that require in-plane confinement of carriers and to explore the low-dimensional physics including quantum point contact and quantum dot. Recently developments in the fabrication of electronic nanostructures has led to the exploration of field-effect devices based on 2D semiconductors such as phosphorene[20,21] and TMDCs.[22–25] In particular, laterally gate-defined quantum devices were realized using molybdenum disulfide ($MoS_2$) heterostructures with high carrier mobility.[26–29] However, high quality quantum confinement in other 2D semiconductors has not been explored to date since the surface of InSe can react with water and oxygen like in many other 2D materials, In particular, oxidation[30,31] can degrade the surface and cause contact barriers[32] leading to highly resistive and non-ohmic contacts, and an underestimation of the field effect mobility.[33,34] Due to the fact that InSe degrades in ambient conditions, the fabrication of high quality InSe device is challenging and the realization of in-plane confinement is so far experimentally unexplored. To prevent the oxidation, few-layer InSe flakes can be encapsulated between hexagonal boron nitride (hBN) layers in a glovebox filled with high purity argon gas.[21,35]

In this manuscript, we report on the fabrication of a gate-defined InSe heterostructure and its electrical properties. To create reliable contacts, we fabricate gates covering regions of overlapping InSe and graphene layers. We find that applying appropriate contact gate voltages helps to prevent the formation of Schottky contacts and to increase the local density of states in this region, leading to linear current-voltage $I$-$V_{sd}$ characteristics, i.e. ohmic behavior. We further utilize top-gates covering InSe in a split-gate geometry to deplete the electron gas in the InSe layer below the gates thus forming a constriction. For high magnetic fields, we observe magnetoresistance minima at even filling factors, which can be accounted for by a two-fold spin degeneracy.

**Methods**

To fabricate our devices we employ the dry transfer technique.[25,27,28] First, we use the top hBN layer (35nm thick) on polymethyl methacrylate (PMMA) membrane to pick up four few-layer graphene flakes serving as electrical contacts to the InSe crystal. Then InSe crystal is mechanically exfoliated on a PPC-covered wafer and identified by optical contrast in the glovebox. A selected six-layer InSe flake is transferred onto the center of the top hBN/graphene heterostructure so that each graphene layer touches the InSe on four sides, as shown in Figs. 1a and 1b. Finally, the heterostructure is deposited onto another hBN flake (8nm thick) on degenerately doped silicon wafers with a 285 nm $SiO_2$ layer. Both top and bottom hBN isolate and protect InSe from the environment and reduce scattering of carriers by phonons and charged impurities in the Si-substrate. The top hBN serves as a dielectric for the split-gates (SG) with a gap of 120 nm and the contact gates (CG) in Fig. 1c. After the assembling process, the resulting heterostructure is taken out of the glovebox. The split-gates and the contact-gates on the top hBN are fabricated by standard electron-beam lithography and metal deposition. Finally, one-dimensional Cr/Au (5/45 nm) edge contacts form the graphene electrodes.[36]

**Results and discussion**

The quality of the contacts is assessed by low temperature ($T = 1.8$ K) two-terminal $I$-$V_{sd}$ characteristics at different $V_{bg}$ and $V_{cg}$, as shown in Fig. 2a. At $V_{cg} = 0$ V and $V_{bg} = 30$, 40 and 50 V, we observe non-linear $I$-$V_{sd}$ curves indicating non-ohmic contacts. However, linear behavior is achieved at $V_{cg} = 16$ V leading to three-fold current enhancement through the InSe channel as compared to $V_{cg} = 0$ V, thus suggesting that the contact gate controls the chemical potential of graphene and tunes the Schottky barrier between the 2D semiconductor and few-layer graphene.[1,23,37] This can be better seen in Fig. 2b where the isolated contact resistances $R_c = \frac{1}{2}(R_{12,12} - R_{12,34})$ are plotted as a function of $V_{bg}$ for various $V_{cg}$ (where $R_{12,12}$ and $R_{12,34}$ are the two-terminal and four-terminal resistances, respectively). For the definition of $R_{PQ,RS}$, the current is injected through contact P and extracted through contact Q. The voltage is measured between contacts R and S. The contact resistance $R_c$ decreases by up to three orders of magnitude with increasing $V_{bg}$ and $V_{cg}$. For $V_{cg} = 12$V and $V_{bg} > 30$V, $R_c$ is sufficiently small (<100 kΩ) for performing four-probe measurements using standard lock-in techniques.

For the determination of the sheet resistivity, we measured the resistance in two configurations, $R_{34,12}$ and $R_{23,41}$, at $V_{cg} = 16$ V, as shown in Fig. 2c. Because the separation between contacts 1 and 4 is shorter than that between contacts 1 and 2, $R_{23,41}$ is smaller than $R_{34,12}$ for all $V_{bg}$. Although the sample does not have an ideal van der Pauw geometry, we roughly estimate the sheet resistivity ($\rho$) of the InSe using the van der Pauw formula with the results shown in Fig 2c. The resistivity varies in the range 1 - 6 kΩ at $V_{bg} > 30$ V and $V_{cg} = 16$ V. At Vbg < 30 V, owing to both the high channel and contact resistances, four-probe measurements using standard lock-in techniques are not possible but only two probe measurements could be carried out. Fig. 2d displays the mobility ($\mu$), as derived from $\mu=1/ne\rho$, as a function of $V_{bg}$, where $n$ is the electron density. The mobility increases with increasing $n$. It reaches a value of ~1,500 cm$^2$/Vs at the maximum $V_{bg}$, smaller than the field effect mobility ( ~2,300 cm$^2$/Vs), which is derived from the $V_{bg}$ dependence of $1/\rho$ in the range $V_{bg}$ = 60-100 V, as shown in the inset of Fig. 2d. This discrepancy is likely to originate from the large size of the contacts compared to the sample area. To investigate the magnetoresistance, we measured $R_{12,43}$ as a function of the magnetic field ($B$) and $V_{bg}$. Although some radial lines at high $B$ are only weakly visible in Fig. 3a, the derivative of $R_{12,43}$ with respect to $B$ clearly shows a Landau fan at $B > 5$ T (Fig. 3b), which suggests a quantum mobility of ~2,000 cm$^2$/Vs in agreement with the field effect mobility. Each dashed line in Fig. 3b corresponds to $v = nh/Be$, where $v$ is the filling factor. The charge carrier density as a function $V_{bg}$ is determined from the SdH (Shubnikov-de Haas) oscillations, as shown in the inset of Fig. 2c. The slope of the density versus gate voltage curve yields a back-gate capacitance $C_{bg}/e = 7.3 \times 10^{10}$ cm$^{-2}$ V$^{-1}$, which is in good agreement with the value $7.36 \times 10^{10}$ cm$^{-2}$ V$^{-1}$ estimated from geometrical considerations taking into account the thicknesses of the SiO$_2$ (285nm) and the bottom hBN (8nm) layers. Shubnikov-de Haas minima match well with even filling factors $v = 14, 16, 18$, which is evidence for the two-fold spin degeneracy expected for InSe.[1]

In the next step, we investigate transport through a split-gate defined constriction with a lithographic gap of 120 nm. First, we apply a positive back gate voltage to create a uniform electron density in the sample. Then negative split-gate voltages are tuned such as to locally deplete the two-dimensional electron gas (2DEG) under the top gates. Fig. 3c shows the conductance ($G_{12,43}$) as a function of $V_{bg}$ and $V_{sg}$ at $V_{cg} = 16$ V. As expected, for higher $V_{bg}$, a larger negative $V_{sg}$ is required to pinch off the channel. The slope of the diagonal line between the red region (highly conductive) and the blue region (nearly insulating) is 8.6, corresponding to a capacitive coupling ratio $C_{tg}/C_{bg} = 8.58$ in agreement with geometrical considerations. As shown in Fig. 3d, for $V_{bg} < 80$ V, the conductance abruptly decreases with increasing $V_{sg}$ towards negative values, indicating that the electron gas below the split-gate begins to deplete. Under these conditions the device becomes insulating with conductance values $G < 0.01$ µS. Interestingly, at $V_{bg} > 80$ V, the drop of the conductance at $V_{sg} \sim -6$V is followed by a slower decrease of the conductance with decreasing $V_{sg}$ and the appearance of reproducible features. This indicates that in the high carrier density regime, after pinching off below the split-gates, it is possible to form a one-dimensional channel. This can be seen from the fact that the channel carrier density ($n_{ch}$) between the split gates can be reduced by $V_{sg}$. Consequently, this transition occurs when $n_{ch}$ reaches a value of ~$2.5 \times 10^{12}$ cm$^{-2}$ and the device is highly resistive, as shown in Fig. 2c and inset. Due to the short mean free path ( <50m) relative to the channel length ( ~120nm), quantized conductance through the one-dimensional channel is not discernible.

To investigate the transport in the 1D channel, we measure the two-terminal AC current ($I_{ac}$) as a function of $V_{sd}$ and $V_{Rsg}$ at $B = 0$ T and $V_{Lsg} = -10$ V (see Fig. 1c for the gate labels). A series of different diamond-shaped zero-current regions appear at $V_{Rsg} < -8$ V, matching the resonance peak positions in Fig. 4b. From the extent of these diamonds in $V_{sd}$-direction, we estimate the charging energy of the localized states, which ranges between 2 and 5 meV. These features are likely to originate from the disordered potential of the InSe layer and require further investigation.

**Conclusion**

In summary, a high quality InSe van der Waals heterostructure encapsulated by hBN are achieved by a novel fabrication process under the oxygen- and moisture-free environment. We have demonstrated two-dimensional and one-dimensional electron transport with an electron mobility of ~2,300 cm$^2$/Vs at 1.8 K by additional electrostatic gating at the contacts and split-top gates. We observe a Landau fan in Quantum Hall regime indicating two-fold spin degeneracy and formation of a narrow conducting channel between split-gates with transport facilitated by localized states. As an electrostatically tunable tunneling barrier reaches complete pinch-off, atomically thin InSe with the relative small in-plane electron mass and high mobility can be an excellent platform to study the low-dimensional physics including quantum dots. Further development of this technology paves the way toward quantized transport and the manipulation of the single charge carriers in the electrostatic confinement.

**Acknowledgments**

We thank Peter Märki and Erwin Studer, as well as the FIRST staff for their technical support. We also acknowledge financial support from the European Graphene Flagship, the Swiss National Science Foundation via NCCR Quantum Science and Technology (QSIT) and ETH Zűrich via the ETH fellowship program, the Engineering and Physical Sciences Research Council [grant number EP/M012700/1] (EPSRC), and the National Academy of Sciences of Ukraine. Growth of hexagonal boron nitride crystals was supported by the Elemental Strategy Initiative conducted by the MEXT, Japan and the CREST (JPMJCR15F3), JST.

**Figure 1. InSe van der Waals heterostructure**. (a) Cross-sectional schematic of an InSe-based field effect device. (b) Optical microscope image of a device. Red and black dashed lines outline the InSe flake and graphene contacts, respectively. The scale bar is 5 μm. (c) Atomic force microscope image of the split-gate ($V_{Lsg}$ and $V_{Rsg}$) with a gap of 120 nm. The scale bar is 500 nm.

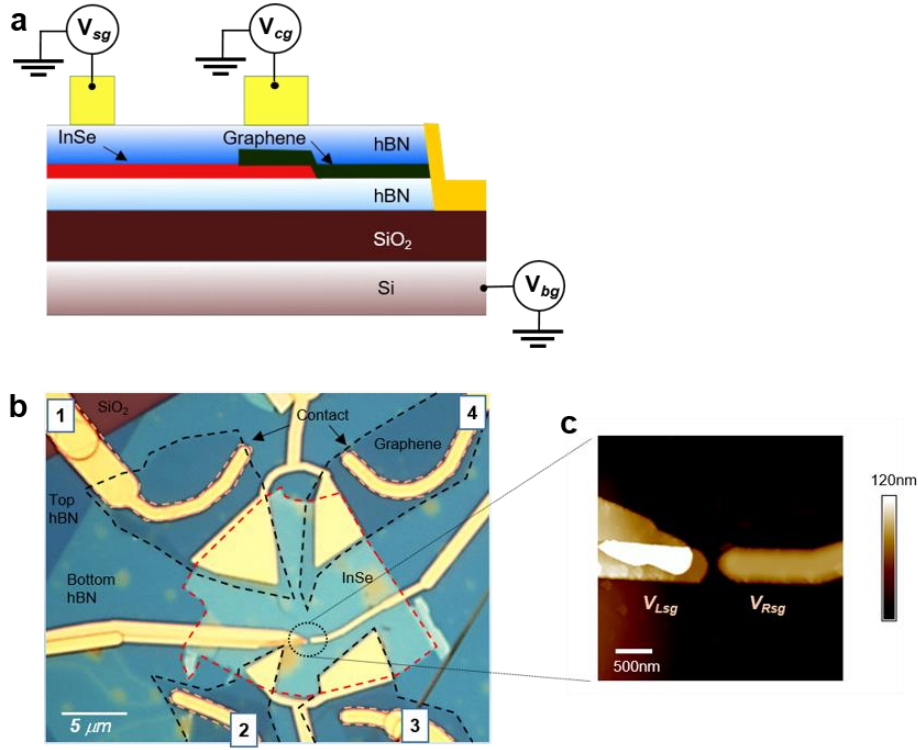

**Figure 2. Transport properties of few-layer InSe at $T = 1.8K$** (a) $I$-$V_{sd}$ characteristic at $V_{bg} = 30$, 40 and 50V when $V_{cg} = 0$ and 16V. (b) Contact resistance as a function of $V_{bg}$ at $V_{cg} = 0$, 6 and 12V. (c) Resistivity as a function of $V_{bg}$ by using the van der Pauw method at $V_{cg} = 16V$. Inset: Carrier density as a function of $V_{bg}$ determined from SdH. (d) Hall mobility $\mu$ as a function of $V_{bg}$. Inset: The conductivity as a function of $V_{bg}$. The linear fitting (orange) is performed to extract the field-effect mobility, ~2,300 cm$^2$/Vs.

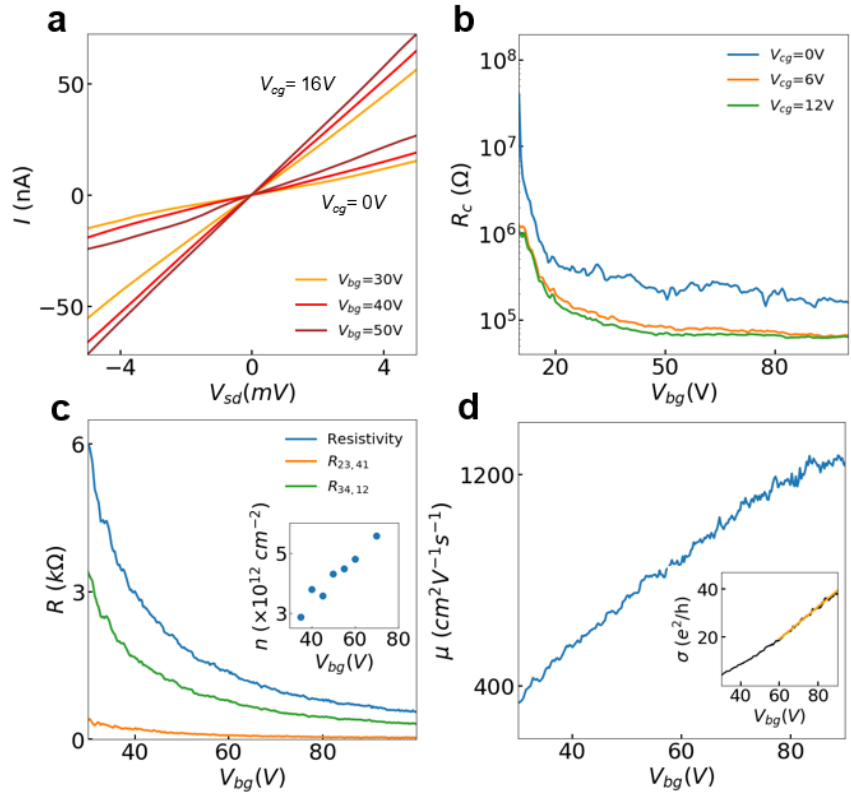

**Figure 3. Magnetotransport and constriction transport in few-layer InSe.**
(a) Resistance and (b) differential resistance as a function of $B$ and $V_{bg}$ at constant $V_{cg}$ = 16V and $T$ = 1.8K. Each dashed line represents an even filling factor. (c) Conductance as a function of $V_{bg}$ and $V_{sg}$. (d) Line traces from (c) at $V_{bg}$ = 60, 70, 80, 90 and 100V.

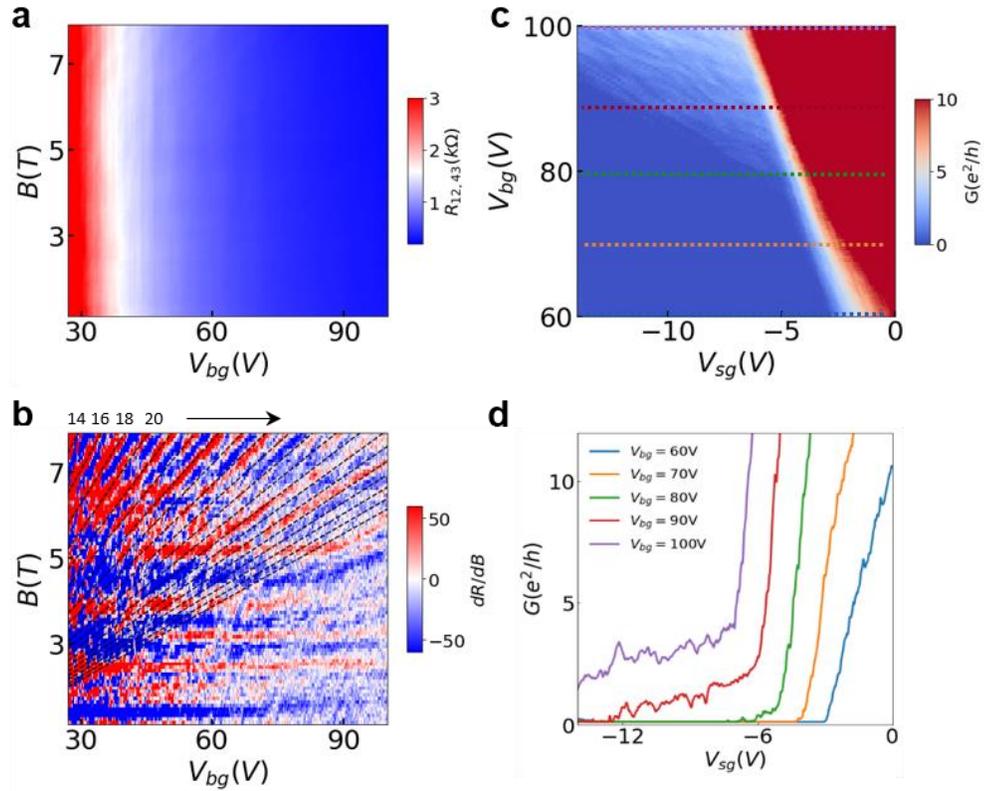

**Figure 4**. (a) Two-terminal pinch off curves as a function $V_{Rsg}$ at $V_{bg} = 82.5V$ and $V_{Lsg} = 10V$. Due to localized states, several resonance peaks appear close to pinch off. (b) Measured AC current as a function of $V_{sd}$ and $V_{Rsg}$ with a 250 µV AC excitation.

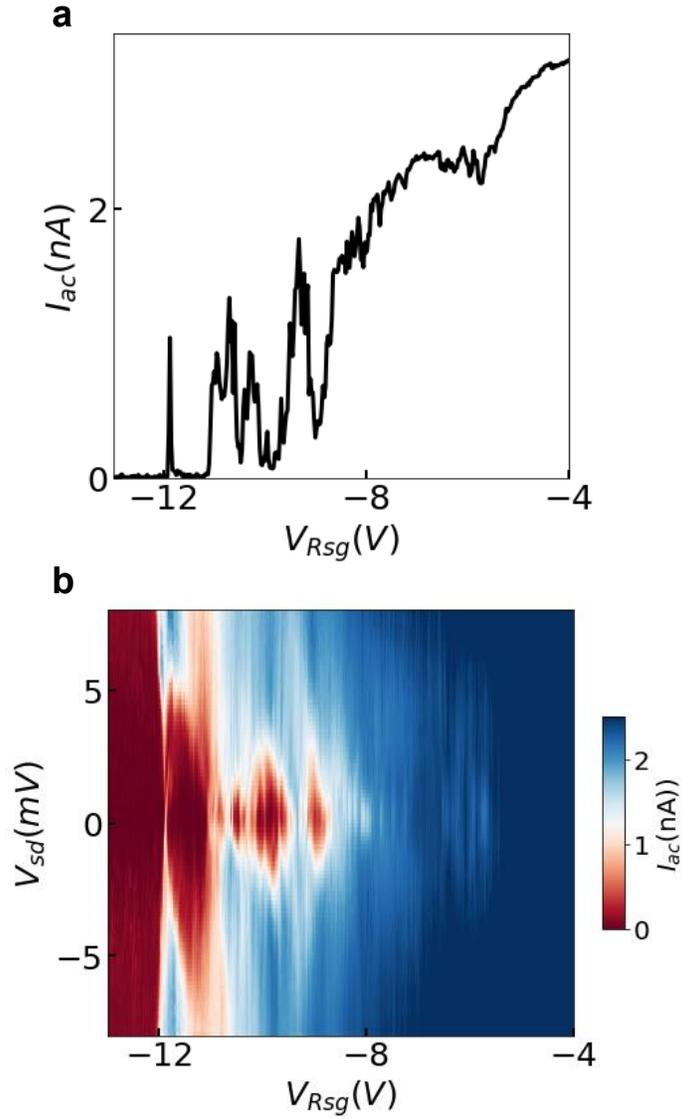